\documentclass[preprints,article,accept,moreauthors,pdftex]{Definitions/mdpi}
\firstpage{1} 
\pubvolume{xx}
\issuenum{}
\articlenumber{}
\pubyear{2021}
\copyrightyear{2021}

\history{}

\usepackage{amsmath}
\usepackage{amsthm}
\usepackage{tikzfig}
\usepackage{amsfonts}

\tikzstyle{env}=[copoint,regular polygon rotate=0,minimum width=0.2cm, fill=black]

\tikzstyle{every picture}=[baseline=-0.25em]
\tikzstyle{dotpic}=[scale=0.5]
\tikzstyle{diredges}=[every to/.style={diredge}]
\tikzstyle{dot graph}=[shorten <=-0.1mm,shorten >=-0.1mm,scale=0.6]
\tikzstyle{plot point}=[circle,fill=black,minimum width=2mm,inner sep=0]
\tikzstyle{linedoted}=\draw[dashed]

\tikzstyle{braceedge}=[decorate,decoration={brace,amplitude=2mm,raise=-1mm}]
\tikzstyle{small braceedge}=[decorate,decoration={brace,amplitude=1mm,raise=-1mm}]
\tikzstyle{left hook arrow}=[left hook-latex]
\tikzstyle{right hook arrow}=[right hook-latex]

\tikzstyle{dtriangle}=[fill=yellow,draw=black,shape=isosceles triangle,shape border rotate=-90,isosceles triangle stretches=true,inner sep=0.8pt,minimum width=0.25cm,minimum height=2mm]
\tikzstyle{vtriang}=[fill=yellow,draw=black,shape=isosceles triangle,shape border rotate=180,isosceles triangle stretches=true,inner sep=0.8pt,minimum width=0.25cm,minimum height=2mm]
\tikzstyle{vrt}=[fill=yellow,draw=black,shape=isosceles triangle,shape border rotate=0,isosceles triangle stretches=true,inner sep=0.8pt,minimum width=0.25cm,minimum height=2mm]
\tikzstyle{H box}=[rectangle,fill=yellow,draw=black,xscale=0.8,yscale=0.8, inner sep=0.6pt]
\tikzstyle{gbox}=[rectangle,fill=green,draw=black,xscale=1.0,yscale=1.0, inner sep=1.pt]
\tikzstyle{rbox}=[rectangle,fill=red,draw=black,xscale=1.0,yscale=1.0, inner sep=1.pt]
\tikzstyle{triangle}=[fill=yellow,draw=black,shape=isosceles triangle,shape border rotate=90,isosceles triangle stretches=true,inner sep=0.8pt,minimum width=0.25cm,minimum height=2mm]

\tikzstyle{bn}=[circle,fill=black,draw=black,scale=.8]
\tikzstyle{wn}=[circle,fill=white,draw=black,scale=.6]
\tikzstyle{dn}=[circle,fill=none,draw=gray]
\tikzstyle{bspider}=[fill=black,draw=black,scale=1,shape=isosceles triangle,shape border rotate=-90,isosceles triangle stretches=true,inner sep=1pt,minimum width=0.4cm,minimum height=3mm]
\tikzstyle{dbspider}=[fill=black,draw=black,scale=1,shape=isosceles triangle,shape border rotate=90,isosceles triangle stretches=true,inner sep=1pt,minimum width=0.4cm,minimum height=3mm]
\tikzstyle{L}=[rectangle,shape=rectangle,fill=green,draw=black]

\tikzstyle{Z dot}=[inner sep=0mm, minimum size=2mm, shape=circle, draw=black, fill={rgb,255: red,221; green,255; blue,221}]
\tikzstyle{Z phase dot}=[minimum size=5mm, font={\footnotesize\boldmath}, shape=rectangle, rounded corners=2mm, inner sep=0.2mm, outer sep=-2mm, scale=0.8, draw=black, fill={rgb,255: red,221; green,255; blue,221}]
\tikzstyle{X dot}=[Z dot, shape=circle, draw=black, fill={rgb,255: red,255; green,136; blue,136}]
\tikzstyle{X phase dot}=[Z phase dot, fill={rgb,255: red,255; green,136; blue,136}, font={\footnotesize\boldmath}]
\tikzstyle{hadamard edge}=[-, dashed, dash pattern=on 2pt off 0.5pt, thick, draw={rgb,255: red,68; green,136; blue,255}]

\tikzstyle{Box1}=[rectangle,shape=rectangle,fill=gray,draw=black,minimum width=0.3cm,minimum height=3mm]

\tikzstyle{dtriangle2}=[fill=gray,draw=black,shape=isosceles triangle,shape border rotate=-90,isosceles triangle stretches=true,inner sep=0.8pt,minimum width=0.25cm,minimum height=2mm]

\tikzstyle{dtriangle3}=[fill=white,draw=black,shape=isosceles triangle,shape border rotate=-90,isosceles triangle stretches=true,inner sep=0.8pt,minimum width=0.25cm,minimum height=2mm]

\tikzstyle{square box2}=[rectangle,fill=white,draw=black,minimum height=5mm,minimum width=10mm,font=\small]

\tikzstyle{scalar}=[rectangle,shape=diamond,fill=white,draw=black, inner sep=0.8pt]

\tikzstyle{black dot}=[inner sep=0.7mm,minimum width=0pt,minimum height=0pt,fill=black,draw=black,shape=circle]

\tikzstyle{dot}=[black dot]
\tikzstyle{smalldot}=[inner sep=0.4mm,minimum width=0pt,minimum height=0pt,fill=black,draw=black,shape=circle]
\tikzstyle{white dot}=[dot,fill=white]
\tikzstyle{antipode}=[white dot,inner sep=0.3mm,font=\footnotesize]
\tikzstyle{smallwhitedot}=[smalldot,fill=white]
\tikzstyle{alt white dot}=[white dot,label={[xshift=3.07mm,yshift=-0.05mm,font=\footnotesize]left:$*$}]
\tikzstyle{gray dot}=[dot,fill=gray!40!white]
\tikzstyle{smallgraydot}=[smalldot,fill=gray!40!white]
\tikzstyle{box vertex}=[draw=black,rectangle]
\tikzstyle{small box}=[box vertex,fill=white]
\tikzstyle{whitebg}=[fill=white,inner sep=2pt]
\tikzstyle{graph state vertex}=[sg vertex,fill=black]

\tikzstyle{wide copoint}=[fill=white,draw=black,shape=isosceles triangle,shape border rotate=90,isosceles triangle stretches=true,inner sep=1pt,minimum width=1.5cm,minimum height=5mm]
\tikzstyle{wide point}=[fill=white,draw=black,shape=isosceles triangle,shape border rotate=-90,isosceles triangle stretches=true,inner sep=1pt,minimum width=1.5cm,minimum height=4mm]
\tikzstyle{very wide copoint}=[fill=white,draw=black,shape=isosceles triangle,shape border rotate=-90,isosceles triangle stretches=true,inner sep=1pt,minimum width=2.5cm,minimum height=4mm]
\tikzstyle{very wide empty copoint}=[draw=black,shape=isosceles triangle,shape border rotate=-90,isosceles triangle stretches=true,inner sep=1pt,minimum width=2.5cm,minimum height=4mm]
\tikzstyle{symm}=[ultra thick,shorten <=-1mm,shorten >=-1mm]

\tikzstyle{square box}=[rectangle,fill=white,draw=black,minimum height=5mm,minimum width=5mm,font=\small]
\tikzstyle{square gray box}=[rectangle,fill=gray!30,draw=black,minimum height=6mm,minimum width=6mm]
\tikzstyle{copoint}=[regular polygon,regular polygon sides=3,draw=black,scale=0.75,inner sep=-0.5pt,minimum width=7mm,fill=white]
\tikzstyle{point}=[regular polygon,regular polygon sides=3,draw=black,scale=0.75,inner sep=-0.5pt,minimum width=7mm,fill=white,regular polygon rotate=180]
\tikzstyle{gray point}=[point,fill=gray!40!white]
\tikzstyle{gray copoint}=[copoint,fill=gray!40!white]

\newcommand{\edgearrow}{{\arrow[black]{>}}}
\newcommand{\edgetick}{{\arrow[black,scale=0.7,very thick]{|}}}

\tikzstyle{diredge}=[->]
\tikzstyle{rdiredge}=[<-]
\tikzstyle{medium diredge}=[->]

\tikzstyle{short diredge}=[->]
\tikzstyle{halfedge}=[-)]
\tikzstyle{other halfedge}=[(-]
\tikzstyle{freeedge}=[(-)]
\tikzstyle{white edge}=[line width=5pt,white]
\tikzstyle{tick}=[postaction=decorate,decoration={markings, mark=at position 0.5 with \edgetick}]
\tikzstyle{small map edge}=[|-latex, gray!60!blue, shorten <=0.9mm, shorten >=0.5mm]
\tikzstyle{thick dashed edge}=[very thick,dashed,gray!40]
\tikzstyle{map edge}=[|-latex,very thick, gray!40, shorten <=1mm, shorten >=0.5mm]
\tikzstyle{tickedge}=[postaction=decorate,
  decoration={markings, mark=at position 0.5 with \edgetick}]
\tikzstyle{dirtickedge}=[postaction=decorate,
  decoration={markings, mark=at position 0.5 with \edgetick},
  decoration={markings, mark=at position 0.85 with \edgearrow}]
\tikzstyle{dirdoubletickedge}=[postaction=decorate,
  decoration={markings, mark=at position 0.4 with \edgetick},
  decoration={markings, mark=at position 0.6 with \edgetick},
  decoration={markings, mark=at position 0.85 with \edgearrow}]

\makeatletter
\newcommand{\boxshape}[3]{%
\pgfdeclareshape{#1}{
\inheritsavedanchors[from=rectangle] 
\inheritanchorborder[from=rectangle]
\inheritanchor[from=rectangle]{center}
\inheritanchor[from=rectangle]{north}
\inheritanchor[from=rectangle]{south}
\inheritanchor[from=rectangle]{west}
\inheritanchor[from=rectangle]{east}
\backgroundpath{
\southwest \pgf@xa=\pgf@x \pgf@ya=\pgf@y
\northeast \pgf@xb=\pgf@x \pgf@yb=\pgf@y

\@tempdima=#2
\@tempdimb=#3

\pgfpathmoveto{\pgfpoint{\pgf@xa - 5pt + \@tempdima}{\pgf@ya}}
\pgfpathlineto{\pgfpoint{\pgf@xa - 5pt - \@tempdima}{\pgf@yb}}
\pgfpathlineto{\pgfpoint{\pgf@xb + 5pt + \@tempdimb}{\pgf@yb}}
\pgfpathlineto{\pgfpoint{\pgf@xb + 5pt - \@tempdimb}{\pgf@ya}}
\pgfpathlineto{\pgfpoint{\pgf@xa - 5pt + \@tempdima}{\pgf@ya}}
\pgfpathclose
}
}}

\boxshape{NEbox}{0pt}{8pt}
\boxshape{SEbox}{0pt}{-8pt}
\boxshape{NWbox}{8pt}{0pt}
\boxshape{SWbox}{-8pt}{0pt}
\makeatother

\tikzstyle{map}=[draw,shape=NEbox,inner sep=7pt]
\tikzstyle{mapdag}=[draw,shape=SEbox,inner sep=7pt]
\tikzstyle{maptrans}=[draw,shape=SWbox,inner sep=7pt]
\tikzstyle{mapconj}=[draw,shape=NWbox,inner sep=7pt]

\tikzstyle{probs}=[shape=semicircle,fill=gray!40!white,draw=black,shape border rotate=180,minimum width=1.2cm]

\tikzstyle{arrs}=[-latex,font=\small,auto]
\tikzstyle{arrow plain}=[arrs]
\tikzstyle{arrow dashed}=[dashed,arrs]
\tikzstyle{arrow bold}=[very thick,arrs]
\tikzstyle{arrow hide}=[draw=white!0,-]
\tikzstyle{arrow reverse}=[latex-]
\tikzstyle{cdnode}=[]

\tikzstyle{gn}=[dot,fill=green,minimum width=0.25cm,inner sep=0pt]
\tikzstyle{rn}=[dot,fill=red,inner sep=0pt,minimum width=0.25cm]

\tikzstyle{rc}=[dot,thick,fill=white,draw = red,minimum width=0.3cm,inner sep=0pt]
\tikzstyle{gc}=[dot,thick,fill=white,draw= green,inner sep=0pt,minimum width=0.3cm]
\tikzstyle{bc}=[dot,thick,fill=white,draw= blue,minimum width=0.3cm]

\tikzstyle{label}=[circle,fill=white,minimum width=0.3cm]

\tikzstyle{clocklabel}=[dot,fill=yellow,draw=black,font=\tiny,inner sep=0.75pt]

\tikzstyle{rsn}=[circle split,draw,fill=red,font=\tiny,inner sep=0.75pt]
\tikzstyle{gsn}=[circle split,draw,fill=green,font=\tiny,inner sep=0.75pt]
\tikzstyle{bsn}=[circle split,draw,fill=blue,font=\tiny,inner sep=0.75pt]

\tikzstyle{rsc}=[circle split,thick,draw= red,draw,fill=white,font=\tiny,inner sep=0.75pt]
\tikzstyle{gsc}=[circle split,thick,draw= green,draw,fill=white,font=\tiny,inner sep=0.75pt]
\tikzstyle{bsc}=[circle split,thick,draw= blue,draw,fill=white,font=\tiny,inner sep=0.75pt]

\tikzstyle{cnot}=[fill=white,shape=circle,inner sep=-1.4pt]
\tikzstyle{wire label}=[font=\tiny, auto]

\tikzstyle{cdiag}=[matrix of math nodes, row sep=3em, column sep=3em, text height=1.5ex, text depth=0.25ex,inner sep=0.5em]
\tikzstyle{arrow above}=[transform canvas={yshift=0.5ex}]
\tikzstyle{arrow below}=[transform canvas={yshift=-0.5ex}]

\usepackage{placeins}
\usepackage{mathtools}
\usepackage{graphicx}
\usepackage{pgfplots}
\pgfplotsset{compat=1.6}
\usepackage[normalem]{ulem}
\usepackage{stmaryrd}
\newtheorem{Postulate}{Postulate}

\Title{Reasoning about conscious experience with axiomatic and graphical mathematics}

\usepackage{color}
\def\bR{\begin{color}{red}}
\def\bB{\begin{color}{blue}}
\def\bM{\begin{color}{magenta}}
\def\bC{\begin{color}{cyan}}
\def\bW{\begin{color}{white}}
\def\bMl{\begin{color}{black}}
\def\bG{\begin{color}{green}}
\def\bY{\begin{color}{yellow}}
\def\e{\end{color}}

\Author{Camilo Miguel Signorelli $^{1,2,*}$, Quanlong Wang $^{1,3}$,Bob Coecke $^{1,3}$}
\AuthorNames{Camilo Miguel Signorelli and Quanlong Wang and Bob Coecke}

\address{%
$^{1}$ \quad Department of Computer Science, University of Oxford\\
$^{2}$ \quad Cognitive Neuroimaging Unit, INSERM U992, NeuroSpin\\
$^{3}$ \quad Cambridge Quantum Computing Ltd.}

\corres{Correspondence: cam.signorelli@cs.ox.ac.uk} 

\abstract{We cast aspects of consciousness in axiomatic mathematical terms, using the graphical calculus of general process theories (a.k.a~symmetric monoidal categories and Frobenius algebras therein). This calculus exploits the ontological neutrality of process theories. A toy example using the axiomatic calculus is given to show the power of this approach, recovering other aspects of conscious experience, such as external and internal subjective distinction, privacy or unreadability of personal subjective experience, and phenomenal unity, one of the main issues for scientific studies of consciousness. In fact, these features naturally arise from the compositional nature of axiomatic calculus.}


\begin{document}
\section{Introduction}   

The main motivation for our theoretical approach is giving formal tools to study consciousness in a rigorous axiomatic setup. Current scientific approaches have thrown away the subjective features of experience, leaving us in a strange position, without rigorous tools to describe qualitative aspects of reality that we experience every day \cite{Goff2019}. To understand this, consider the next common example: if a tree falls and nobody is there to hear it, does the tree make any sound? Yes, of course, the tree generates vibrations, but the quality of sounds are only assigned by the observer. In other words, there are objective realities (vibrations), but subjective and qualitative features such as sounds, colours, smells and tastes exist  only if a conscious mind is ready to experience them. The vibration is characterized by common mathematical language and physical mechanisms, while qualitative and subjective aspects do not have any formal mathematical language to refer to them. 

We claim here that axiomatic reasoning in the form of  graphical calculi  (a.k.a.~compositional mathematics)  may bring the uniqueness of conscious experience back to science, constructing a new form of describing the structure of experience from its direct phenomenology \cite{Merleau-Ponty2005,Husserl1983} and therefore, a new science of consciousness.  

Graphical calculi  naturally arise in  category theory \cite{MacLane1998}, specifically symmetric monoidal categories  \cite{Coecke2011a}, also called process theories \cite{Coecke2017a}. Because of their abstract mathematical nature, they also are ontologically neutral, i.e. processes in a theory do not assume any concrete physical realization. One can extend this idea to mental processes without any lack of generality. Therefore, it is equally valid to suggest an interpretation of graphical calculus starting from mental processes than from physical ones. It makes process theories and graphical calculus optimal setups to explore the assumption of consciousness and subjectivity as fundamental processes of nature. Here, such fundamental processes are modelled by the irreducible and primitive nature of the mathematical generators on that calculus.   

Briefly, this article explores a re-interpretation of process theory and graphical calculi in the context of formal structures of conscious experience  \cite{Yoshimi2007,Prentner2019,Tsuchiya2020} and process philosophy \cite{Rescher,Whitehead1929}. This approach attempts to model what \textit{consciousness itself is doing}, instead of what the brain or any other physical system is \textit{doing} regarding conscious experience. The mathematical formalism of process theories is first introduced and motivated by concrete examples (Section \ref{secc:CSCE}). Then, the definition of conscious experience is constructed via entangled features of that experience, its phenomenology and empirical distinctions (Section \ref{secc:FC}). These definitions are interpreted as mathematical generators to provide a reasoning example, from which a more complex property of conscious experience arises: namely the the structure of privacy or unreadability of others personal experiences (Section \ref{secc:PC}). Moreover, we address, restate and discuss the question about the unity of consciousness according to compositional approaches (Section \ref{secc:PU}). Finally, we conclude with how the use of process theories and axiomatic mathematics brings new advantages in the formal study of conscious experience (Section \ref{secc:Disc} and \ref{secc:Con}), in line with the contemporary research direction of mathematical consciousness science \cite{AMCS2021} and phenomenology \cite{Merleau-Ponty2005,Thompson2007}.

\section{Pictorial mathematics for conscious experience}\label{secc:CSCE}

Across this section, we introduce the mathematical formalism of process theories. 

\subsection{Process theories}\label{secc:Processtheories}

The formalism of \emph{process theories} \cite{Coecke2017a} provides  a graphical language to reason about processes as abstract mathematical entities. These graphical languages are based on symmetric monoidal categories  \cite{Coecke2011a}, making them mathematically rigorous frameworks. The main components are \emph{systems}, or more generally speaking, \emph{types}, which are   represented  by wires (e.g.~type $A$ and type $B$), and \emph{processes},  represented by boxes with a number of input and output wires, which vary from box to box in number and labelling.  In short, processes correspond to transformation.  Some diagrammatic examples are:

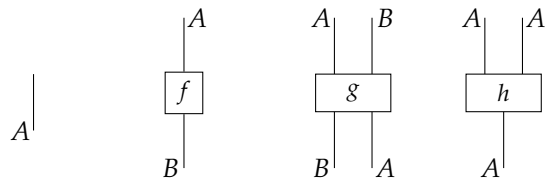
\begin{center}
\beginpgfgraphicnamed{TikZit//Bob1}
\InputIfFileExists{TikZit//Bob1.tikz}{}{\input{./figures/TikZit//Bob1.tikz}}
\endpgfgraphicnamed
\end{center}

Reading the diagrams from top to bottom,\footnote{Note that in much of the literature on process theories, the wires are read in the opposite direction, bottom-up.} the process $f$ 
can be thought of as a  map $f:A\rightarrow B$ from $A$ to $B$, the process $g$  as a  map $g:A\otimes B\rightarrow B\otimes A$, and $h$  as  $h: A\otimes A \rightarrow A$,  where the symbol `$\otimes$' stands for `composing systems'.   The symbol `$I$', stands for `no system', which, evidently, is graphically represented by `no wire'.  This induces  special processes, such as \emph{states}, \emph{tests} and \emph{numbers}, with associated maps   $\phi:I\rightarrow A$, $\varphi:A\rightarrow I $, and $s: I\rightarrow I$ respectively. Graphically, they are represented as follows:

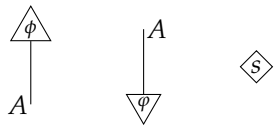
\begin{center}
\beginpgfgraphicnamed{TikZit//Bob2}
\InputIfFileExists{TikZit//Bob2.tikz}{}{\input{./figures/TikZit//Bob2.tikz}}
\endpgfgraphicnamed
\end{center}

These graphical forms have the power to make simple the reasoning about how systems and processes interact in different contexts. In this line, what is important in process theories is how systems and processes compose. Basically, there are two main types of composition, the sequential composition given by $\circ$ and the parallel composition described by $\otimes$\footnote{This symbol is indeed used both for composing systems and processes.}. If two processes $f$ and $g$ interact (and their wires match), these two compositions look like:

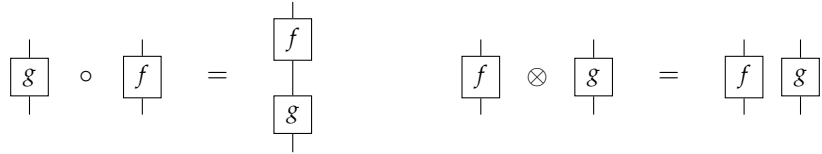
\begin{center}
\beginpgfgraphicnamed{TikZit//Bob3}
\InputIfFileExists{TikZit//Bob3.tikz}{}{\input{./figures/TikZit//Bob3.tikz}}
\endpgfgraphicnamed
\end{center}

Usually, these diagrams are represented and restricted to one dimensional expression, namely $(f\circ g$) and $(f\otimes g)$ respectively. Interestingly, the graphical two dimensional representation allows us to move the process boxes up and down freely, helping us to prove equalities and making reasoning about processes much more intuitive, for example: 

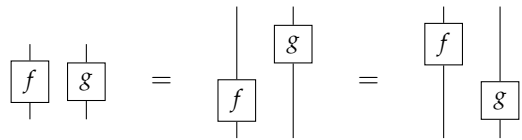
\begin{center}
\beginpgfgraphicnamed{TikZit//Bob4}
\InputIfFileExists{TikZit//Bob4.tikz}{}{\input{./figures/TikZit//Bob4.tikz}}
\endpgfgraphicnamed
\end{center}

The advantage, thereof, is a very intuitive notion of equality: processes are equal when they are represented by the same diagram. The two dimensional graphical representation also let us prove more complex equations in simple forms, making them almost tautological. For instance, we leave to the reader the task of drawing the diagrams for the following one-dimensional equation and check by themselves that the equation easily holds in graphical form:

\begin{equation*}
 (f \otimes g) \circ (h \otimes k) =  (f \circ h) \otimes (g \circ k)  
\end{equation*}

From this graphical notion of equality  there emerges a second more precise one: two diagrams are equal when one becomes the other via certain transformations. Interestingly, these transformations are purely topological, or more accurately, these transformations follow the principle that only connectedness matters. It means that we can obtain certain results by looking at the relevant diagrams, since these results are already present in the topology of the corresponding graph. This concept is reviewed with an example in next subsections.  

\subsection{Interpretation of a theory}

All process theories share one  mathematical structure, i.e.~the structure of symmetric monoidal categories. Using category theoretic terminology, a \emph{functor}  is a map $F: \textbf{M} \rightarrow \textbf{N}$ that   translates a process theory \textbf{M} into another  process theory \textbf{N}. In other words, $F$ assigns each system $A$ in \textbf{M} to a system $FA$ in \textbf{N}, and each process $f: A \rightarrow B$ in \textbf{M}, to a process $Ff: FA \rightarrow FB$ in \textbf{N}, obeying certain equations ensuring that sequential and parallel compositions are respected \cite{Awodey2006,Coecke2016}. This functor can also be understood as an \textit{interpretation} of the theory \textbf{M} in \textbf{N}. When working with diagrams, to specify  an interpretation it is enough to specify the images of the diagrams.  Of course, there could exist many such interpretations. The image of \textbf{M} in \textbf{N} is called a \textit{model}.

\subsection{Generators and rewriting rules} \label{Gen}

Process theories enable one to axiomatise theories  in a variety of disciplines, and may reveal that theories from very different scientific areas may share a surprising amount of common structure \cite{Signorelli2020a}. A striking example is the structural commonality of quantum theory and natural language \cite{Coecke2013}. Also here we will encounter a similar remarkable structural correspondence.    

A specific process theory  may be characterised by a generating set of systems and processes.  General systems and processes are then obtained by composing these, that is, by making the generators interact.  It may be the case that there is only one generating 
system and very few generating processes. Conceptually, we think of these generators as basic (or primitive) systems and processes. 

The full specification of a process theory, in addition to the presentation of the generators, then also tells us what these generators stand for.  To better understand this, consider the following example. The next four diagrams are the basic maps or transformations in a process theory of Boolean Circuits. 

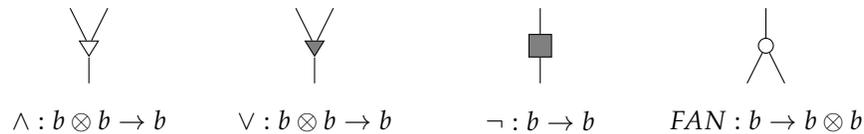
\begin{center}
%
\beginpgfgraphicnamed{TikZit//Bob5b}
\InputIfFileExists{TikZit//Bob5b.tikz}{}{\input{./figures/TikZit//Bob5b.tikz}}
\endpgfgraphicnamed
\end{center}

The first operation corresponds to the logic gate \textit{and}, the second one to \textit{or}, then \textit{negation} and \textit{FAN} operation, respectively. In this theory, the basic system is the bit $b$, given by the pair of values $B(b)=\{0,1\}$. These values come from the chosen interpretation for these diagrams, the specific mapping $B: \textbf{BoolCirc} \rightarrow \textbf{Bool}$. This mapping translates the above diagrams into a concrete calculus:\newline

$ B(\wedge)= a:\left\{\begin{array}{ccc}00 \mapsto 0\\ 01 \mapsto 0\\ 10 \mapsto 0\\ 11\mapsto 1  \end{array} \right.$     $B(\vee)= o:\left\{\begin{array}{ccc}00 \mapsto 0\\ 01 \mapsto 1\\ 10 \mapsto 1\\ 11 \mapsto 1  \end{array} \right. $ $B(\neg)= n:\left\{\begin{array}{ccc}0 \mapsto 1\\ 1 \mapsto 0 \end{array} \right.$   $B(FAN)= \delta: \left\{\begin{array}{ccc}0 \mapsto 00\\ 1 \mapsto 11 \end{array} \right.$ \newline

With these four generators, a more complex process in this theory is represented as the composition of these generators. For example, the logical expression: $(x \wedge \neg y) \vee \neg (y \wedge z)$, becomes a circuit of logic gates such as the result depends on the state value entered into the circuit and the interpretation of the generators given above. Graphically:

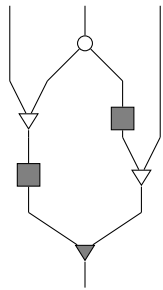
\begin{center}
%
\beginpgfgraphicnamed{TikZit//Bob6b}
\InputIfFileExists{TikZit//Bob6b.tikz}{}{\input{./figures/TikZit//Bob6b.tikz}}
\endpgfgraphicnamed
\end{center}

The functor $B$ is one possible mapping, but there are other alternatives as well. This presentation of generators is the basic syntax of a theory. In order to capture the full theory, we need some extra equations on the class of diagrams. These equations are called \textit{rewriting rules}. These rules are basically a pair of diagrams of the same type that correspond to an equivalence or equality between each other. For example, the next composition of a $\vee$ and $\wedge$ is rewritten as the composition of $FAN$, $\vee$ and $\wedge$: 
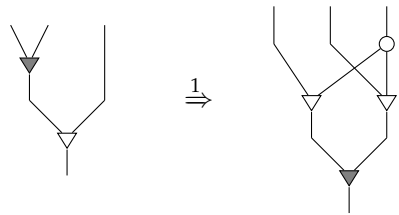
\begin{center}
%
\beginpgfgraphicnamed{TikZit//Bob7b}
\InputIfFileExists{TikZit//Bob7b.tikz}{}{\input{./figures/TikZit//Bob7b.tikz}}
\endpgfgraphicnamed
\end{center}
Careful reading shows that this rule corresponds to the distributive law. Another example is to rewrite the diagram composed by sequentially connecting $\wedge$ and $\neg$, using the equivalent diagram formed by two $\neg$ and one $\vee$:
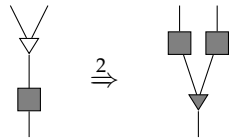
\begin{center}
%
\beginpgfgraphicnamed{TikZit//Bob8b}
\InputIfFileExists{TikZit//Bob8b.tikz}{}{\input{./figures/TikZit//Bob8b.tikz}}
\endpgfgraphicnamed
\end{center}

Together these two rules allow us to rewrite as follows, where rule 2 is applied first, followed by rule 1:
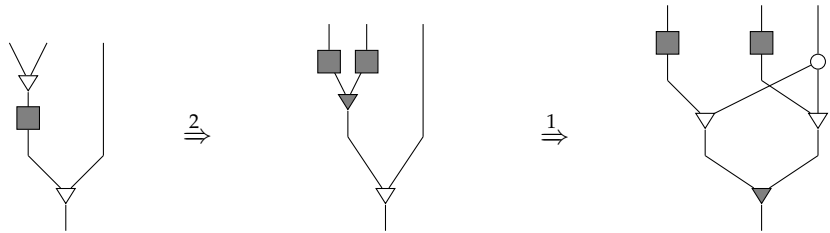
\begin{center}
%
\beginpgfgraphicnamed{TikZit//Bob9b}
\InputIfFileExists{TikZit//Bob9b.tikz}{}{\input{./figures/TikZit//Bob9b.tikz}}
\endpgfgraphicnamed
\end{center}

Here we will always assume that if there is a rewriting rule, $ d \xRightarrow{\gamma} d'$,  there is also a rewriting rule $ d' \xRightarrow{\gamma'} d$.  Then, these rewriting rules lead us to a formal definition of equality across diagrams. 

\begin{Definition}\label{equality}
Given a set of rewriting rules $\Gamma$, a diagram $d$ and another $d'$ are considered equal $d = d'$, if via applying rewriting rules in $\Gamma$, $d$ becomes $d'$. We denote the existence of such a rewrite as $ d \xRightarrow{\Gamma} d'$.  
\end{Definition}

For an interpretation $F: \textbf{M} \rightarrow \textbf{N}$, thinking of \textbf{M} as diagrams and rewrites with equality as defined above, \em soundness \em means that $ d \stackrel{\Gamma}{=} d'$ implies $ Fd \stackrel{\Gamma}{=} Fd'$.  If  $ Fd = Fd'$ moreover also implies $d \stackrel{\Gamma}{=} d'$, the interpretation is called \textit{complete}. In other words, no more equalities hold for the model than can be derived by the rewriting rules. 

\subsection{Process theory and consciousness}

We shortly account for using process theory as a mathematical framework for a theory of conscious experience. As reviewed above, process theories present various advantages: i) they are rigorous and intuitive reasoning tools, ii) they focus on processes composition instead of objects, iii) they admit simple ways to prove complex equations, iv) they define equality in intuitive topological terms, and v) they generalize and compare theories from first principles (axioms). 
Can those properties bring new insights into studies of conscious experience?

In this new context, we can exploit the features of process theories under the philosophical umbrella of its ontological neutrality. Diagrams come with no ontology. Indeed, their ontology only arise in relationship with what is expected to be described, via defining a functor. Process theories deal with the phenomenon and do not claim anything about fixed properties such as mass or charge. On the contrary, everything that exists is studied as a process of changes and transformations \cite{Rescher,Whitehead1929}. These features place process theories in a phenomenological and pragmatic ground that allow us to study the experience from axioms directly obtained by the experience itself. In other words, process theories licence us to suspend the query of ontological discussions while reinforcing an epistemic caution \cite{Varela1996,Lusthaus2002}: the world appears to us only in co-relationship with us, becoming specified through sense-making.  

Although we highlight the \textit{primacy} of conscious experience, we leave aside the deeper epistemic and ontological interpretations about that claim. Here, we focus only on the pragmatic aspect. We propose a compositional model with experiential processes as generators and interpret a set of rewriting rules as compositions and modifications of experiences. These rules specify the generators via allowed relationships (compositions) \cite{Signorelli2020c}, becoming more concrete instances of experiences.

In the following, this paper does not attempt to give any complete interpretation $F$, between the hypothetical category of conscious experiences, let's say \textbf{CExp}, and a graphical calculus in the symmetric monoidal category  \textbf{Mon}, such as $F: \textbf{CExp} \rightarrow \textbf{Mon}$. 
Instead, we show how a fragment of the ZW-calculus \cite{Hadzihasanovic2018}, whose diagrams compose a symmetric monoidal category, enables us to perform formal reasoning about conscious experience.

\section{Defining generators for conscious experience}\label{secc:FC}

In order to define mathematical generators for conscious experience, we may introduce as few  assumptions as possible. The introduction of a few concepts \cite{Searle2000,Lusthaus2002,Merleau-Ponty2005,Block2005,Bayne2012}, however, seems enough to recover properties of consciousness within our formal model (Section \ref{secc:PC}). In this section we first present semi-formal statements about conscious experience as motivations to find a formal counterpart within graphical calculi.

\subsection{A phenomenological hypothesis}

Our main assumption departs from current axiomatic studies of physical theories. Usually, we map the phenomenon under consideration into one specific category, via functor definition. Here, we assume that the diagrams of symmetric monoidal category already conveys the basic phenomenology of our experience. For instance, sequential compositions may involve phenomenological aspects of internal time-consciousness in Husserl’s discussions \cite{Husserl1964}. According to phenomenological interpretations \cite{Merleau-Ponty2005, Husserl1964}, the structure of symmetric monoidal categories might already reflect the structure of experience. Then, theoretical axiomatizations in the field of physics may also accept a reinterpretation as mapping  physical phenomena (e.g. classical mechanics, quantum mechanics, relativity, etc) into the structure of our conscious experience. Of course, we need much more work to formalize and align this assumption with Husserl’s and modern phenomenology \cite{Yoshimi2007}. 

Under this hypothesis, we define a type $A$ of a symmetric monoidal category as a primary/minimal undefined or indistinguishable experience. We also introduce a process called the identity $1_A$, which does nothing at all to A. 

$$ %
\beginpgfgraphicnamed{TikZit//SystemAsolo}
\begin{tikzpicture}
	\begin{pgfonlayer}{nodelayer}
		\node [style=none] (8) at (0, 0.25) {};
		\node [style=none] (9) at (0, -0.5) {$ A \quad$};
	\end{pgfonlayer}
	\begin{pgfonlayer}{edgelayer}
		\draw (8.center) to (9.center);
	\end{pgfonlayer}
\end{tikzpicture}
}
\endpgfgraphicnamed $$

Then, morphisms become transformations (as dimensions of  experience), that are themselves also experiences. For instance, the symmetry condition of symmetric monoidal category is given by a \textit{swap} experience, such that: 
$$%
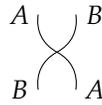
\beginpgfgraphicnamed{TikZit//swap2v}
\InputIfFileExists{TikZit//swap2v.tikz}{}{\input{./figures/TikZit//swap2v.tikz}}
\endpgfgraphicnamed$$

We can also define a notion of experiences that `invert' experiences, i.e. they introduce duals (e.g. opposite relations such as above and below). We call those processes caps $\eta_{A} : I \rightarrow  A^{*} \otimes A$ and cups $\epsilon_{A} : A \otimes A^{*} \rightarrow I$,  respectively signified by: 

$$%
\beginpgfgraphicnamed{TikZit//cap}
\begin{tikzpicture} [scale=0.7]
	\begin{pgfonlayer}{nodelayer}
		\node [style=none] (0) at (0.5, -0.25) {$A$};
		\node [style=none] (1) at (-0.5, -0.25) {$A^{*}$};
	\end{pgfonlayer}
	\begin{pgfonlayer}{edgelayer}
		\draw [in=90, out=90, looseness=1.75] (1.center) to (0.center);
	\end{pgfonlayer}
\end{tikzpicture}}
\endpgfgraphicnamed \quad \quad %
\beginpgfgraphicnamed{TikZit//cup}
\begin{tikzpicture} [scale=0.8]
	\begin{pgfonlayer}{nodelayer}
		\node [style=none] (0) at (0.5, 0.25) {$A^{*}$};
		\node [style=none] (1) at (-0.5, 0.25) {$A$};
	\end{pgfonlayer}
	\begin{pgfonlayer}{edgelayer}
		\draw [in=-90, out=-90, looseness=1.75] (1.center) to (0.center);
	\end{pgfonlayer}
\end{tikzpicture}}
\endpgfgraphicnamed $$

It adds a compact close structure \cite{Selinger2011,Signorelli2020c}.

\subsection{Unity}

Unity of experience is one of the most salient features of consciousness as a natural process \cite{Prentner2019}. Any experience is given as a unified single moment and seems irreducible. Some may argue this experience is continuous, others that it is discrete \cite{VanRullen2003, Wittmann2011}, it may contain one or many different contents, etc. Independently, the subsumed experience is one unified coexistence, a unified conscious field \cite{Searle2000} that may be just conceptually subdivided into different notions of unity (Objectual, Spatial, Subjective, Subsumptive) \cite{Bayne2012}. 

In compositional models, unity is realized by non-trivial composition of different processes.  Unity is an intrinsic property of processes, such that a process would `possess' unity as long as it cannot be written as a disconnected diagram. One example of this non-trivial composition corresponds to entangled states in quantum theory. The entanglement is modelled by the use of caps and cups that allow us to relate the notions of sequential and parallel composition: 

$$  %
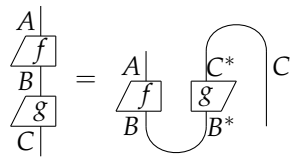
\beginpgfgraphicnamed{TikZit//timespace2}
\InputIfFileExists{TikZit//timespace2.tikz}{}{\input{./figures/TikZit//timespace2.tikz}}
\endpgfgraphicnamed $$

In other words, an experience that comes before another ($f$ before $g$), is equivalent to the experience $f$ happening simultaneously with the inverse of the experience $g$, as far as they reorganize via caps, cups, swaps and/or identity. If this is the case, we said both experiences ($f$ and $g$) are unified in one single experience. 

\begin{Definition}
Unity of experience is realized by non-trivial composition, such that an experience process \textit{possess} unity as long as it cannot be written as a disconnected diagram.
\end{Definition}

For conceptual convenience, we represent unity by the four mathematical diagrams: cap, cup, swap and identity, as examples of basic unity processes and different forms of experiential unity, but also because they allow us to reorganize and compound other processes/experiences into complex diagrams that will remain connected. By consequence, we can  manipulate diagrams to accommodate and visualize their compositions. 

In order to make the following arguments simpler, we will use a self-dual structure \cite{Selinger2011,Signorelli2020c}.

\subsection{Qualitative and subjective processes}\label{QandSprocesses}

Another important feature of conscious experience is that it involves a \textit{qualitative} dimension. Every experience is mostly qualitative, rather than quantitative \cite{Goff2019}. In the words of Nagel, there is a kind of "it feels like" or "what is it like to be" something or someone having certain experience \cite{ThomasNagel1974}. The qualitative character of experience may come from external perceptions or internal thoughts \cite{Searle2000}. Indistinctly, both are unified experiences involving qualitative descriptions that cannot be easily measured. These descriptions are what distinguishes between the experience of red and green: the irreducible phenomenology of consciousness, phenomenal consciousness, minimal phenomenal experience, or qualia \cite{Metzinger2020a,Block2005}. 

Category theory and its graphical forms allow us a very intuitive first approximation to formally describe this qualitative dimension. In algebra, common operations, such as addition ($+$) and multiplication ($\times$), follow axioms like associativity and commutativity. For an arbitrary operation ($\star$) and elements $a$, $b$, and $c$, associativity looks like:

\begin{equation}\label{product}
\beginpgfgraphicnamed{TikZit//Associativity22}
\InputIfFileExists{TikZit//Associativity22.tikz}{}{\input{./figures/TikZit//Associativity22.tikz}}
\endpgfgraphicnamed
\end{equation}

This algebraic structures, together with its unit %
\beginpgfgraphicnamed{TikZit//unit}
\begin{tikzpicture}[scale=0.5]
	\begin{pgfonlayer}{nodelayer}
		\node [style=wn] (0) at (0, 0.25) {};
		\node [style=none] (1) at (0, -0.25) {};
	\end{pgfonlayer}
	\begin{pgfonlayer}{edgelayer}
		\draw (0) to (1.center);
	\end{pgfonlayer}
\end{tikzpicture}
}
\endpgfgraphicnamed, is called a \textit{monoid} \footnote{A monoid is always a pair of diagrams, i.e. the two legs white node (e.g. multiplication) and the state (unit).}. The graphical form contains a topological \textit{intuition} behind the notion of associativity \cite{Lawvere2009}. This notion is qualitative, it is not quantifiable per se, as the reader may observe from the equation above. In other words, these diagrams may carry some \textit{qualitative} structure of formal statements.  
Additionally, we can also conceptualize quantity as a form of quality, a very precise, unfuzzy one. As such, quality may subsume quantity, making qualitative aspects more general than quantitative ones. In the above example, the quantitative aspect is realized by the specification of the operation $\star$ and the elements $a$, $b$, and $c$.

\begin{Definition}\label{qualia}
Qualitative structure of experience is represented by diagrammatic equations, such that an experience process \textit{possess} quality as long as it contains trivial topological relationships that are non-trivial for formal statements.
\end{Definition}

A more general algebraic structure, the Frobenius algebra is built by the monoid and its \textit{comonoid} \footnote{A comonoid is also a pair of diagrams, i.e. the copy-like node (e.g. comultiplication) and the effect/test (counit).}. The comonoid is basically the same diagram than above, but inverted, such that monoid and comonoid follows the next Frobenius law. 
$$  %
\beginpgfgraphicnamed{TikZit//Frobenius}
\InputIfFileExists{TikZit//Frobenius.tikz}{}{\input{./figures/TikZit//Frobenius.tikz}}
\endpgfgraphicnamed $$

If we add an extra condition,

$$  %
\beginpgfgraphicnamed{TikZit//specialcondition}
\InputIfFileExists{TikZit//specialcondition.tikz}{}{\input{./figures/TikZit//specialcondition.tikz}}
\endpgfgraphicnamed $$

this is called a Special Frobenius algebra. We can condense such structure, symmetric and commutative conditions within an abstract mathematical entity called white spider (where all white dots merge together for arbitrary number of elements).

\begin{equation}\label{phasewhitespiderd}
\beginpgfgraphicnamed{TikZit//phasewhitespider}
\InputIfFileExists{TikZit//phasewhitespider.tikz}{}{\input{./figures/TikZit//phasewhitespider.tikz}}
\endpgfgraphicnamed
\end{equation}

This spider seems topologically trivial, but contains not-trivial algebraic structure. Following the preliminary definition \ref{qualia} and the intuition that quality subsumes quantity, the white spider will be called a qualitative process. In our framework, the qualitative structure of experience is denoted by this unspecified process, where $r\in R$ is a parameter taking values in an arbitrary commutative ring, associated with quantitative aspects that qualitative experience may also carry. Please note that this spider is slightly different than the example in equation \ref{product}, since dots and legs denote different operations and types of elements. Moreover, the qualitative process from ZW-calculus \cite{Hadzihasanovic2015,Hadzihasanovic2018} can be generalized to the Z (green) spider with multiple parameters as given in \cite{Signorelli2020c, Wang2021}. 

We can further postulate that the composition for qualitative processes correspond to the next rule:

\begin{Postulate}\label{Qualia}
Qualitative process compounds as follows: 
\begin{equation}\label{eqQualia}
\beginpgfgraphicnamed{TikZit//whitespiderfusion}
\InputIfFileExists{TikZit//whitespiderfusion.tikz}{}{\input{./figures/TikZit//whitespiderfusion.tikz}}
\endpgfgraphicnamed
\end{equation}
where $rs$ is the  product of $r$ and $s$.
\end{Postulate}

It means that any qualitative aspect of the experience (given by the process of quality) is fused and glued by default, just by means of being connected. Note that this is a non-trivial consequence of associativity and the Frobenius conditions introduced above.

Additionally, any conscious experience has also a  \textit{subjective} dimension, perhaps, inseparable of the qualitative one \cite{Searle2000}. It seems that experiences only exist if there are subjects or agents (sentient beings) to experience something. Neither does a rock appears to have any kind of experience, nor particles or atoms. Qualitative processes would imply subjective ones since, for a qualitative feeling regarding some
event to exist, there must exist a subject to experience that event \cite{Searle2000}. This experience is part of the so-called first-person accounts, corresponding to elements of reality that do not exist without a subject, such as perceptual experiences (e.g. the experience of colour), bodily experiences (e.g. pain and hunger), emotional experiences, mental imagery, among others \cite{Chalmers2013}. First-person accounts contrast with the third-person accounts, related to "objective" and quantitative measurements such as brain signatures of perceptual discrimination or differences between sleep and wakefulness \cite{Chalmers1995}. Therefore, conscious experiences seems to exist only when there are agents to experience: some “I” owner of that experience. This imposes a boundary that perceived elements must "cross" to become part of that experience. This is called conscious access.  

In order to account for this intrinsic relationship between qualitative and subjective dimension of experience, we tentatively define a subjective process using an adaptation of the mathematical comonoid introduced in ZW-calculus \cite{Hadzihasanovic2018}, represented by a black triangle.

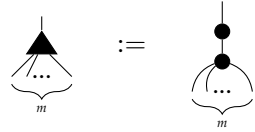
\[ %
\beginpgfgraphicnamed{TikZit//zstm}
\InputIfFileExists{TikZit//zstm.tikz}{}{\input{./figures/TikZit//zstm.tikz}}
\endpgfgraphicnamed\]where $m\geq 2$.

 The white monoid and its black comonoid follow the \textit{bialgebra} law.

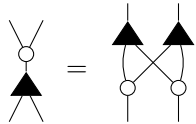
\[ %
\beginpgfgraphicnamed{TikZit//bialgebralaw}
\InputIfFileExists{TikZit//bialgebralaw.tikz}{}{\input{./figures/TikZit//bialgebralaw.tikz}}
\endpgfgraphicnamed\]

The formal definition of this graphical form is standard in the literature and detailed discussions can be found in \cite{Selinger2011} and \cite{Coecke2011}, among many others.

\begin{Definition}
Qualitative and subjective dimensions of experience are realized by a bialgebra structure, such that a qualitative process is the monoid and subjective one is the comonoid, each one forming a Special Frobenious algebra.
\end{Definition}

In short, the subjective process is a generalization of the triangle %
\beginpgfgraphicnamed{TikZit//trianglesingle}
\InputIfFileExists{TikZit//trianglesingle.tikz}{}{\input{./figures/TikZit//trianglesingle.tikz}}
\endpgfgraphicnamed, and its unit (a.k.a. effect):

$$  %
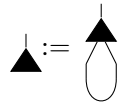
\beginpgfgraphicnamed{TikZit//blackeffect}
\InputIfFileExists{TikZit//blackeffect.tikz}{}{\input{./figures/TikZit//blackeffect.tikz}}
\endpgfgraphicnamed $$

such that we can define its own monoid, states, and identity. 

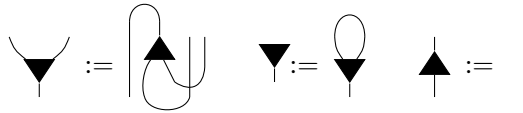
\[%
\beginpgfgraphicnamed{TikZit//downblacktriangle}
\InputIfFileExists{TikZit//downblacktriangle.tikz}{}{\input{./figures/TikZit//downblacktriangle.tikz}}
\endpgfgraphicnamed \quad\quad %
\beginpgfgraphicnamed{TikZit//unitdowanblack}
\InputIfFileExists{TikZit//unitdowanblack.tikz}{}{\input{./figures/TikZit//unitdowanblack.tikz}}
\endpgfgraphicnamed \quad\quad %
\beginpgfgraphicnamed{TikZit//identityblack}
\InputIfFileExists{TikZit//identityblack.tikz}{}{\input{./figures/TikZit//identityblack.tikz}}
\endpgfgraphicnamed \]

Note that these diagrammatic definitions use the caps and cups, while the black triangle with one input and one output coincides with the identity process, all them introduced in previous section. Finally, we can recursively define the black triangle with multiple legs, leading us to the almost tautological second postulate, a similar rule of composition for subjective process.

\begin{Postulate}\label{Subject}
Subjective process compounds as follows: 
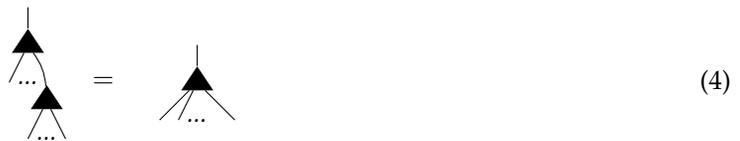
\begin{equation}\label{eqSubject}
\beginpgfgraphicnamed{TikZit//zstassocia}
\InputIfFileExists{TikZit//zstassocia.tikz}{}{\input{./figures/TikZit//zstassocia.tikz}}
\endpgfgraphicnamed
\end{equation}
\end{Postulate}

These composition rules ensure the unity of experience and its compositional nature across different instances of experience. In both cases, the composition takes the form of a fusion rule given by associativity axioms.

\subsection{Distinction}
\label{secc:D}

Qualitative and subjective processes in terms of  
two dimensions of conscious experience  result in  
two different kinds of unities that in turn generate distinctions. We interpret the former as  the \textit{phenomenal unity} and the latter as  the \textit{access unity} \cite{Bayne2012} (a.k.a phenomenal consciousness and access consciousness). The distinctions correspond to distinctive experiences and distinctive content, respectively, the "how" we are conscious and about "what" we are conscious of \cite{Weisberg}. Phenomenal experience, the what is like to be, is differentiated from the access consciousness, i.e. the accessibility of content for further cognitive processing in a certain moment of time \cite{Block2005, Aru2012}. The difference is not only conceptual, but it also seems to involve empirical evidence of different brain signatures \cite{Block2005, Aru2012}. It is important to mention this, because assumptions in our model do correspond to these conceptual but also empirical division. 

In our framework, conscious experience generates distinctions that break the invariability of the primitive unity and creates different ways to discriminate between subject and object, quality and quantity, inside or outside, identical or different, among others. To include this relevant aspect of experiences, we invoke the last attribute called the \textit{distinction} process.  
The distinction applies between experiences but also distinguishing among elements on that experience. Due to a normal form of ZW-calculus \cite{Hadzihasanovic2018}, the distinction diagram can be constructed from qualitative diagrams and subjective diagrams, but for simplicity and convenience, it is represented as another new process.

\begin{Definition}
Distinction is a primary generator, represented by: 
 \[%
\beginpgfgraphicnamed{TikZit//braid}
\InputIfFileExists{TikZit//braid.tikz}{}{\input{./figures/TikZit//braid.tikz}}
\endpgfgraphicnamed\]
\end{Definition}

\section{Composition of conscious experience}
\label{secc:PC}

In this section, we define and implement possible rewriting rules for conscious experience. The set of processes introduced above become the generators of our calculus (Table \ref{generators}), while extra operations and rewriting rules form part of the explicit axioms in the theory. These axioms specify the generators, as discussed in section \ref{Gen}.

\begin{table}[ht]
\begin{center} 
\begin{tabular}{|c|}
\hline
\\
$\left\llbracket \quad %
\beginpgfgraphicnamed{TikZit//Cap}
\InputIfFileExists{TikZit//Cap.tikz}{}{\input{./figures/TikZit//Cap.tikz}}
\endpgfgraphicnamed  \quad %
\beginpgfgraphicnamed{TikZit//Cup}
\InputIfFileExists{TikZit//Cup.tikz}{}{\input{./figures/TikZit//Cup.tikz}}
\endpgfgraphicnamed \quad %
\beginpgfgraphicnamed{TikZit//swap}
\InputIfFileExists{TikZit//swap.tikz}{}{\input{./figures/TikZit//swap.tikz}}
\endpgfgraphicnamed \quad %
\beginpgfgraphicnamed{TikZit//Id}
\InputIfFileExists{TikZit//Id.tikz}{}{\input{./figures/TikZit//Id.tikz}}
\endpgfgraphicnamed \quad\right\rrbracket=\textbf{Unity}$ \\
\\\hline
\\
 $\left\llbracket%
\beginpgfgraphicnamed{TikZit//phasewhitespider}
\InputIfFileExists{TikZit//phasewhitespider.tikz}{}{\input{./figures/TikZit//phasewhitespider.tikz}}
\endpgfgraphicnamed\right\rrbracket=\textbf{Qualitative}$\\
 \\\hline 
 \\
$\left\llbracket%
\beginpgfgraphicnamed{TikZit//trianglewstate}
\InputIfFileExists{TikZit//trianglewstate.tikz}{}{\input{./figures/TikZit//trianglewstate.tikz}}
\endpgfgraphicnamed\right\rrbracket=\textbf{Subjective}$  \\
\\\hline
\\
$\left\llbracket%
\beginpgfgraphicnamed{TikZit//braid}
\InputIfFileExists{TikZit//braid.tikz}{}{\input{./figures/TikZit//braid.tikz}}
\endpgfgraphicnamed\right\rrbracket= \textbf{Distinction}$ \\ 
\\ \hline
\end{tabular} \caption{Generators for a graphical calculus of conscious experience. These processes are taken from ZW-calculus \cite{Hadzihasanovic2018} and their graphical forms naturally arise in monoidal categories \cite{Coecke2011a}.}\label{generators} 
\end{center}
\end{table}

\subsection{The relational nature of experience}\label{everything}

In section \ref{secc:FC}, we introduced provisional definitions and interpretations of the generators. In strict sense, they do not model any phenomena by themselves, but only when they are specified by rewriting rules or relationships between them.

This \textit{relational nature} of graphical calculi is relevant, since experience seems also specified in reference to other experiences \cite{Tsuchiya2020,Signorelli2020c}, and being co-dependent \cite{Signorelli2020}. In this paper, for example, unity of experience conveys relationships between qualitative and subjective dimensions of that unity, namely phenomenal unity and access unity, represented by our white and black generators. Both types of unity-experience create distinctions, the former differentiate among experiences, while the latter among contents of those experiences. Then distinctions are signified by the crossing generator. Relationships between generators lead to more complex process compositions, while their behaviour is assumed here as the minimal structure of experience. Therefore, conscious experience is both: the entangled composition of all these processes, as well as from which those conceptual distinctions arise. 

The rest of this article specifies the role of these generators via relational rewriting rules, reinterprets the behaviours of these generators, and from them infers new features of conscious experience.

Importantly, all the rewriting rules follow mathematical considerations, either from standard bialgebras \cite{Selinger2011, Coecke2011} or from the specificity of ZW-calculus \cite{Hadzihasanovic2018}. However, the particular set of generators and rewriting rules are chosen because they do make “sense” for a theory of conscious experience, and not because they are nice mathematically or fit any physical theory. One example is the structure of privacy or personal experience, as we demonstrate in the following sections.

\subsection{Conscious experience}

Following previous discussions, we can define conscious experience in a rigorous graphical form. It is easily done as a composition of qualitative and subjective processes. Therefore, we postulate:
 
\begin{Postulate}\label{Experience} \textbf{Conscious experience}. 
Conscious experiences correspond to compositions of qualitative and subjective processes, such that the composition generates a new diagram, representing a new kind of experience. 
\end{Postulate}

The allowed compositions are subject to a fixed collection of rewriting rules. In this theory, these rewriting rules might correspond to specific set of experiences.

Let's take a first rule from the symmetric monoidal category of ZW-calculus and reinterpret it as the composition of one input quality process carrying a quantitative value $r$,  and one subjective process with two outputs. This composition generates the experience of copy that quality, and we called it experience 1.

\begin{equation}\label{eq4}
\beginpgfgraphicnamed{TikZit//wphasecopy}
\InputIfFileExists{TikZit//wphasecopy.tikz}{}{\input{./figures/TikZit//wphasecopy.tikz}}
\endpgfgraphicnamed  \quad\quad \mapsto Experience  \quad 1
\end{equation}

As we mention before, the way how to read these diagrams is from top to bottom, i.e. imagine the %
\beginpgfgraphicnamed{TikZit//phaser}
\begin{tikzpicture}
	\begin{pgfonlayer}{nodelayer}
		\node [style=none] (0) at (0, 0.25) {};
		\node [style=none] (1) at (0, -0.25) {};
		\node [style=wn] (2) at (0, 0) {$r$};
	\end{pgfonlayer}
	\begin{pgfonlayer}{edgelayer}
		\draw (0.center) to (1.center);
	\end{pgfonlayer}
\end{tikzpicture}}
\endpgfgraphicnamed "crossing"  %
\beginpgfgraphicnamed{TikZit//trianglesingle}
\InputIfFileExists{TikZit//trianglesingle.tikz}{}{\input{./figures/TikZit//trianglesingle.tikz}}
\endpgfgraphicnamed to modify the original shape of the diagram, as shown by the equality. In this case, a subjective process takes and makes a "copy" of a qualitative one to make it available to other mental operations.

Another rewriting rule is the composition between two inputs, one qualitative process and one subjective process, generating another type of experience in our formal model.  

\begin{equation}
\beginpgfgraphicnamed{TikZit//wphasebialgebra}
\InputIfFileExists{TikZit//wphasebialgebra.tikz}{}{\input{./figures/TikZit//wphasebialgebra.tikz}}
\endpgfgraphicnamed \quad\quad \mapsto Experience  \quad 2
\end{equation}

The only difference between both rules is the number of inputs in the qualitative process. In this axiomatic model, conscious experiences are unified compositions of qualitative and subjective generators related to the shape-effect, or circuit reorganization of the diagrams, generated by rules of composition.

Phenomenologically speaking, these rules are important because they introduce the notion of co-dependency between qualitative and subjective experience. Two processes co-dependent if they are co-defined. For example, the first rule tells us that a qualitative process is an experience that can be copied by a subjective process in order to be experienced. On the other hand, a subjective process is an experience that can copy a qualitative one. Then, the division between qualitative and subjective experience becomes conceptual, since any conscious experience is simultaneously qualitative and subjective. As a consequence, it might not be surprising that current experiments do not entirely dissociate phenomenal and access aspects of conscious experience. It might be an implication of the parallel made between qualitative and subjective processes with the phenomenal versus access consciousness that we have introduced previously.

\subsection{Distinctions and boundaries}

According to the previous definition of conscious experience, an important question concerns  how to distinguish two elements already bound (more details in section \ref{secc:PU}). To target this last question, the distinction process seems to present a compelling property:

\begin{Postulate}\label{Distinction} \textbf{Distinction.}
Distinction differentiates between qualitative and subjective processes as follows: 
\begin{equation}\label{Distinctioneq}
\beginpgfgraphicnamed{TikZit//phasecrossingFull}
\InputIfFileExists{TikZit//phasecrossingFull.tikz}{}{\input{./figures/TikZit//phasecrossingFull.tikz}}
\endpgfgraphicnamed
\end{equation}
\end{Postulate}

In other words,  qualitative aspects of experience are the processes that "cross distinctions", while subjective processes do not. It generates, indeed, a distinction between subjects-objects and between internal-external experiences, since another subject is always external to the observer subject. This notion is formalized by using the distinction process as separator or boundary, becoming one of our postulates:

\begin{Postulate}\label{Boundery1} \textbf{Boundary.}
Distinction generates a boundary between external and internal experiences.
 \begin{equation}\label{Boundery}
\beginpgfgraphicnamed{TikZit//swapboundary}
\InputIfFileExists{TikZit//swapboundary.tikz}{}{\input{./figures/TikZit//swapboundary.tikz}}
\endpgfgraphicnamed 
 \end{equation}
\end{Postulate}

Interestingly, it seems that a basic notion of \textit{conscious agent} \cite{Hoffman2014} could arise from the composition of distinction process as a boundary and subjective processes. In this case, however, the agent has the potential of conscious experience but it does not convey an conscious experience itself, according to our postulate \ref{Experience}. Future works may clarify and extend this implications/interpretations.

Additionally, we might introduce two extra rules. The composition between one subjective two output process and one qualitative two inputs process is interpreted as the creation of one subjective state and one subjective effect (perhaps, understood as voluntary action). Graphically:

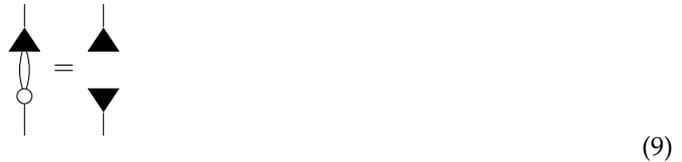
\begin{equation}\label{eq6}
\beginpgfgraphicnamed{TikZit//rule6}
\InputIfFileExists{TikZit//rule6.tikz}{}{\input{./figures/TikZit//rule6.tikz}}
\endpgfgraphicnamed
\end{equation}

The rule above defines two subjective entities from the decoupling of both generators. In other words, if these subjective and qualitative processes compound, such as their outputs and inputs match (always multiple of 2), the result are one subjective state and one subjective effect, like the forms introduced in section \ref{secc:Processtheories} and \ref{QandSprocesses}. 

Another relevant rule is about a distinction process interacting with a subjective two outputs process.

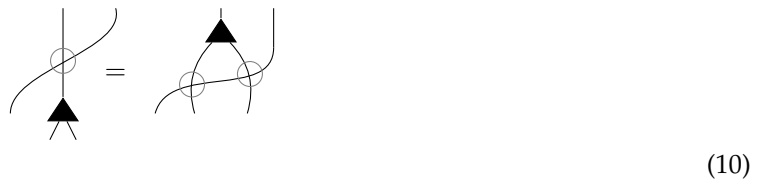
\begin{equation}\label{rule7}
\beginpgfgraphicnamed{TikZit//rule7}
\InputIfFileExists{TikZit//rule7.tikz}{}{\input{./figures/TikZit//rule7.tikz}}
\endpgfgraphicnamed
\end{equation}

Following postulate \ref{Boundery1} and equation \ref{Boundery}, in this case the first subjective process is interpreted as internal and the second as external. Interestingly, this rule informs us about the "generation" of two distinction processes, each time that one distinction process compose with another two outputs internal subjective process, or conversely, the need of two distinctions to transform one external two outputs subjective process into an internal one (see section \ref{secc:PU}).

\subsection{Private experience}
At this point, all the basic compositions and interpretations of the model are in place to define conscious perception as the simple composition of all these generators:

\begin{Postulate}\label{Perception}\textbf{Conscious perception.}
Conscious perception corresponds to the composition of qualitative, subjective and distinction processes, together with its modifications via rewriting rules. 
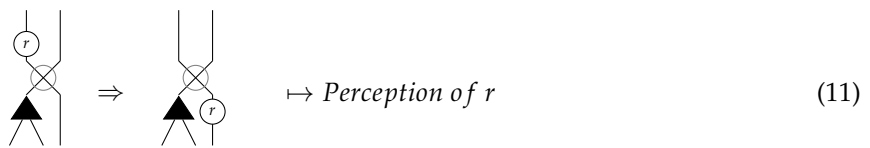
\begin{equation}
\beginpgfgraphicnamed{TikZit//perceptiontry}
\InputIfFileExists{TikZit//perceptiontry.tikz}{}{\input{./figures/TikZit//perceptiontry.tikz}}
\endpgfgraphicnamed \quad\quad \mapsto Perception ~ of ~ r
\end{equation}
\end{Postulate}

Perception is not only a conscious experience, but it is also the kind of conscious experience that generates distinctions between external and internal contents of that experience. This external versus internal division is what is called objective versus subjective division, since an externally triggered experience is associated with objective perception, while internally triggered experiences (what happens after crossing the distinction boundary) are commonly related to subjective inner experiences. However, in our model this objective/subjective division is illusory, every external process is a combination of qualitative and subjective processes, and the properties of the distinction process makes us perceive them differently. In other words, the objective versus subjective divisions is a consequence of our own operation of perceiving. 

When we say “illusion”, it is not in the common sense of “illusionism” in philosophy of mind, where the mind is neglected and considered just as an illusion given by the brain, its neurons and other physical systems. In our case, what is illusory is the distinction between objective world and subjective world. Everything is some kind of “experiential qualitative and subjective world”. However, this “everything is an experience” does not necessarily convey a claim of ontological primacy, since an epistemic primacy suffices to support that claim as well. In this work, we do not commit to any ontological nor epistemic interpretation, but we pragmatically focus on the primacy of experience via the generators, compositions and their consequences.

As a way of example, we can use the postulates above to infer and prove the most salient and recognised property of conscious experience, namely, its private \cite{Searle2000} or better understood personal aspect \cite{Varela1996}. Importantly, this is a pure consequence of the axioms above.

\begin{Proposition}\textbf{Unreadability of others/external subjectivity.}
It is impossible to fully perceive, access or read others'/external conscious subjective experiences. 
\end{Proposition}

\begin{proof} 
From postulates \ref{Experience} and \ref{Perception}, conscious perception involves qualitative, subjective and distinction processes, such that the distinction imposes a boundary between external and internal experiences (equation \ref{Boundery}). Moreover,  equations in \ref{Distinctioneq} force a restriction to subjective processes, preventing them from crossing the boundary. Graphically: \newline

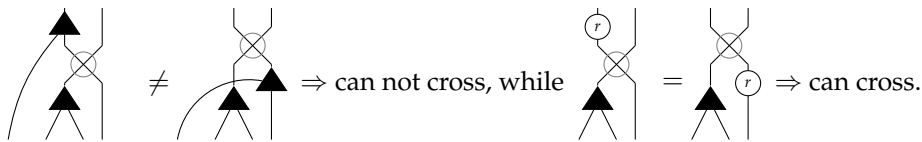
\quad\quad \quad\quad 
\beginpgfgraphicnamed{TikZit//noaccess22}
\InputIfFileExists{TikZit//noaccess22.tikz}{}{\input{./figures/TikZit//noaccess22.tikz}}
\endpgfgraphicnamed $\Rightarrow$ can not cross, 
while  %
\beginpgfgraphicnamed{TikZit//goaccess2}
\InputIfFileExists{TikZit//goaccess2.tikz}{}{\input{./figures/TikZit//goaccess2.tikz}}
\endpgfgraphicnamed $\Rightarrow$ can cross.\newline

It completes the proof. 
\end{proof}

This simple example shows the power of graphical reasoning and axiomatic mathematics to formalise the structure of conscious experience. We have recovered from first principles, one of the main and more recognized hallmarks of personal and subjective conscious experience \cite{ThomasNagel1974,Chalmers1995,Thompson2007}: the inaccessibility of others' subjective "what is like to be" becomes a consequence of a simple and topological property of the graphical calculus introduced. 

This proof, however, does not mean we never have access to others/external subjective experiences. The rule given by equation \ref{rule7} allows us to access those experiences only if we impose more distinctions. In other words, if we have access to it (e.g.~via verbal report), we experience new distinctions that are not present in the original experience. Therefore, as pointed out by \cite{Varela1996}, subjective experience might not be really private, but personal. 

\section{The combination of experiences}\label{secc:PU}

In this section, we explore another insight from our logic approach: the phenomenal unity of experience.

\subsection{The problem of unity}
Among the questions about the structure of conscious experience, how to combine more basic experiences becomes one of the most problematic issues. This is the question about how gluing different elements of experience in one unified phenomenal experience: the unity thesis. In other words, how to combine objects, feelings and other background feature to generate one single unified phenomenal subjective experience \cite{Bayne2012}. 

This problem has two main dimensions, the phenomenal unity and the access unity \cite{Revonsuo1999, Bayne2012}. The former is sometimes called the \textit{combination problem} and the later the \textit{segregation problem}. The first one corresponds to the intuition that regardless of the distinct elements of experiences, they are always integrate-wholes, i.e. the what is like to be in such experience is one whole experience. The second problem is that regardless of distinct and combined features, our experience can segregate elements to recognize different contents of such experience \cite{Treisman1999, Velik2012, Feldman2013}.  These two problems imply the identification of a complete set of fundamental experiences from which other experiences combine, such that the segregation is always regard to this fixed set of experiences \cite{Chalmers2016}.

 \subsection{Phenomenal and access unity}
 
In our compositional model, these questions are stated differently. First, the combination problem does not exist anymore. In fact, it is replaced by a \textit{decomposition problem}. The unity of experience is given by default, just by means of being compositional.  According to our theory, the unity of consciousness, and specifically phenomenal unity, is given by the primary graphical generators and  through topological connection. Secondly, any experience might be always decomposed into combinations of these generators. Which makes the decomposition problem tractable within our formalism.
 
At the same time, the segregation of certain elements of perception is targeted by the distinction process and modifications of compounded qualitative and subjective processes. In other words, the issues become a problem of modification \cite{Searle2000}. Our approach guides us to search for mechanisms of separation and distinction that make elements of our perception look segregated, instead of looking at how to "integrate" or unify elements already unified. 

Graphically, segregation is commonly represented by processes such as:

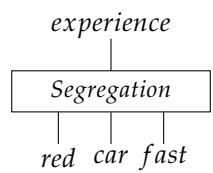
\begin{center}
\beginpgfgraphicnamed{TikZit//Bob10}
\InputIfFileExists{TikZit//Bob10.tikz}{}{\input{./figures/TikZit//Bob10.tikz}}
\endpgfgraphicnamed
\end{center}

In this case, any process theory needs to implement extra processes to account for these decomposition processes, like the decomposition framework introduced in \cite{Tull2020}.   

Slightly different, in our specific model, the question becomes how experiences modify each other to account for distinctions among experiences. It implies that different conscious perceptions are indeed modifications of an already existing field of consciousness, instead of built from various disparate bits of reality \cite{Searle2000}. The main difference is that in cognitive neuroscience the segregation problem is about recognition of "external objects", while here, individual entities arises from unified quality/subjectivity, i.e. the evolution of subjective experiences correspond to modifications or modulations of a unified and already existing qualitative subjectivity, the intrinsic mental consciousness that is independent of the five senses \cite{Llinas1998}. Take, for example, a more complex experiential structure given by the next composition:
 
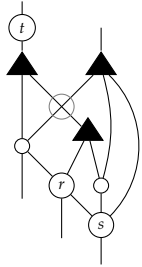
\[%
\beginpgfgraphicnamed{TikZit//Circuit6-3}
\InputIfFileExists{TikZit//Circuit6-3.tikz}{}{\input{./figures/TikZit//Circuit6-3.tikz}}
\endpgfgraphicnamed\]
 
If we use the rewriting rules from equations and examples in section \ref{secc:PC}, we can simplify this circuit as follows:
  
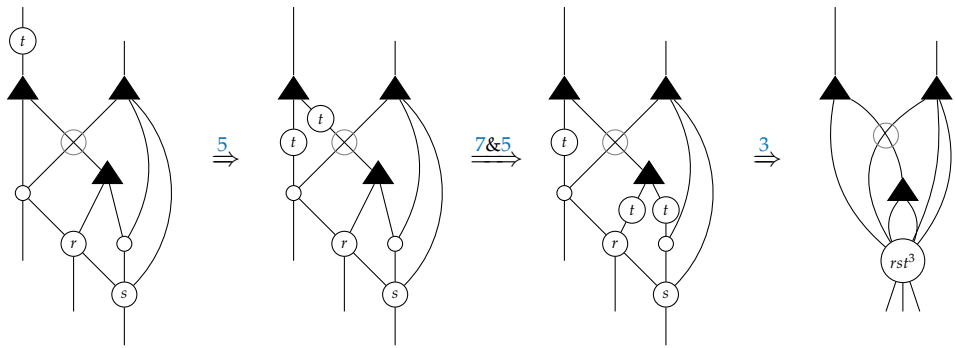
\[%
\beginpgfgraphicnamed{TikZit//Circuit7-2}
\InputIfFileExists{TikZit//Circuit7-2.tikz}{}{\input{./figures/TikZit//Circuit7-2.tikz}}
\endpgfgraphicnamed\]
 
In our model, each shape-effect or modification of the diagrams models a new instance of experience, e.g. a "raw" experience into a thought about it. The circuit itself corresponds to the phenomenal unity or phenomenal field. The most basic conscious experience is the total phenomenal experience. This field is basic but not less complex structure, given by different types of sub-circuits from which distinctions arise. For instance, if we continue applying other rules, we obtain the following circuit: 

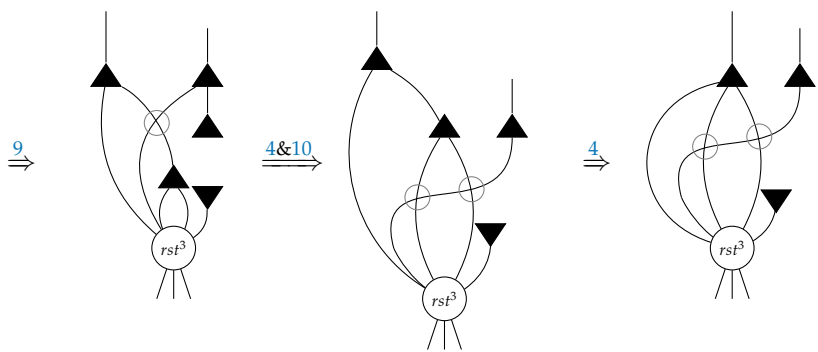
\[%
\beginpgfgraphicnamed{TikZit//Circuit8-2}
\InputIfFileExists{TikZit//Circuit8-2.tikz}{}{\input{./figures/TikZit//Circuit8-2.tikz}}
\endpgfgraphicnamed\]

This diagram can be further simplified (e.g. the right upper triangle becomes an identity, etc). To make our point, however, it is enough that the reader notices how a new distinction process appears. These new distinctions are the effects of the four generators interacting and reorganizing the circuit formed by them. Here, the reorganization that gives rise to new distinctions is what corresponds to the segregation, such that perceptual acts that segregate the content of experience are modelled by the appearance of new distinction processes (rule \ref{rule7}), and its eventual “crossing” (postulate \ref{Perception}). Therefore, given a primary total field, or global consciousness, its modifications through rewriting rules (interpreted as concrete instances of experiences, section \ref{secc:PC}) inform about particular perceptual states, the access unity of contents of experience. These states may represent specific individual loci of quality-subjectivity through the specification of the types/systems. The missing ingredient in this discussion is the empirical translation into phenomenological meaningful patterns for each individual, something that we would expect to implement for each particular case. Although we do not illustrate any particular perceptual act, in the example above, a total phenomenal unity subsumes any other distinctive perception \cite{Bayne2012}. Further research and phenomenological accounts may bring more light on these implications.

To summarise, the phenomenal unity is expressed by primitive notions of unity as generators of our calculus, while feature segregation, as form of access unity, is the disruption or distortion of that phenomenal unity. 

\section{Discussion}
\label{secc:Disc}

Our approach is a provisional proof of concept,  that form part of a new contemporary research direction to study the mathematical structure of consciousness and mathematize phenomenology \cite{Yoshimi2007,Prentner2019,Tsuchiya2020}. The graphical calculus introduced here is one of the mathematical structures to reach that goal. Other examples using different flavours of process theories are \cite{Signorelli2020c} and \cite{Tull2020}. Thus, our model is not the unique model nor the unique way to mathematically study properties of conscious experience.   

Nevertheless, our study of mathematical structures implies a new paradigm to deal with basic assumptions, clearly motivated by phenomenology \cite{Merleau-Ponty2005,Husserl1983} and process philosophy \cite{Rescher,Whitehead1929}. In this article, we started by the assumption of conscious experience as a fundamental process and building the rest of the theory via explicit graphical axioms given by rewriting rules. These axioms are inspired by phenomenological considerations and differently than previous approaches that also claim to start from phenomenological axioms \cite{Oizumi2014}, in our case, all the features of conscious experience have a direct mathematical counterpart. These generators form a minimum set, axioms intend to subsume both phenomenological and mathematical meaning, and nothing extra than what is explicitly stated across these pages is assumed (e.g.~we do not need to assume a physical classical world).

While  process theories are ontologically neutral and our main assumption is based on the hypothesis of the primacy of consciousness, our theory conveys a clear advantage over other models of consciousness: the direct link with physical theories and their mathematical structures. For example, our approach is related almost tautologically to foundations of physics \cite{Coecke2011} and specifically quantum theory. In our model, similar generators and rewriting rules form part of the ZW-calculus \cite{Hadzihasanovic2015}. ZW-calculus was developed for qubits, it is a sound and complete semantic for graphical treatments inside categorical quantum theory \cite{Abramsky2004}, and also a useful graphical language to reconstruct different aspects of physical theories \cite{Coecke2010a,Coecke2011, Hadzihasanovic2015, DeFelice2019}. Without lacking any mathematical rigour, across this article we have reinterpreted the nature of a partial set of their generators and rewriting rules. Accordingly, the connection with fundamental physical theories is reached only invoking phenomenal aspects, no need for any ontological assumption. In other words, it does not matter whether a process is a mental process or a physical process, they share a similar mathematical structure.

We feel that a science of consciousness has much more to gain using high-level mathematical formalisms than focusing on the physical ontologies that may, or may not explain consciousness \cite{Signorelli2021b}. For instance, we do not claim ZW-calculus is complete for a theory of consciousness. Otherwise, this completeness would mean that there is nothing more to consciousness than there is to a qubit. More interesting, however, it is to develop further graphical calculus to search for a complete and sound description based on well informed phenomenological inputs. In other words, directly axiomatize the phenomenology of conscious experience using graphical calculi, and study the models arising from them.

\section{Conclusions} 
\label{secc:Con}

In this article, we introduced a new  paradigm to reason about conscious experience. This graphical interpretation is based on symmetric monoidal categories and follows similar principles and mathematical structures that have proved useful in the foundations of physical theories \cite{Coecke2011}. Moreover, our discussion takes inspiration from the hypothesis of conscious agents \cite{Hoffman2014,Fields2018}, phenomenology \cite{Thompson2007,Merleau-Ponty2005,Signorelli2020}, Buddhist phenomenology \cite{Lusthaus2002, Makeham2014}, as well as the unified field hypothesis \cite{Searle2000} and compositional models \cite{Coecke2013, Coecke2016}. 

Using this compositional framework and primitive mathematical generators as essential features of conscious experience, we recovered different aspects of experience: external and internal subjective distinction, private or personal experience, and phenomenal unity. All of them arise naturally as a consequence of a formal theory of conscious experience that takes the experience as a fundamental process of nature.

In this line, these types of models may become a formal tool to study the phenomenology of cognitive experience in general, and the phenomenology of conscious experience in particular. Philosophers and neuroscientists can also benefit from these intuitive forms \cite{Landry2018,Gomez-ramirez2014,Signorelli2020e}, describing and discussing in graphical terms the  basic assumptions of their respective models \cite{JohnnesKleiner2019a}.

The future for these axiomatic models is promising and exciting. On the one hand, one can extend these descriptions to a better-informed set of generators and rewriting rules. To reach this goal we can use more detailed insights from the phenomenology of experience, micro-phenomenology protocols \cite{Petitmengin2018} and neuro-phenomenology \cite{Varela1996}, as well as contemplative sciences, among others methods. For instance, one may like to define the entire set of axioms and rewriting rules for a sound and complete calculus taking further phenomenological considerations.  On the other hand, one may also expect the objective realm arising from basic experiential generators \cite{Signorelli2020c}. To this end, the goal is recovering objective physical theories from primitive experience that indeed become a mirror of each other \cite{Signorelli2020a}, the very notion of time, probably being one of the most relevant \cite{Husserl1964,Kent2021}. In both research projects, process theories resonate with philosophical phenomenology, avoiding any ontological claim as well as the need for invoking any physical realization but pure mathematical entities.

\vspace{6pt} 



\authorcontributions{Conceptualization, CMS and QW; investigation CMS, QW and BC; writing-original draft preparation, CMS; writing-review and editing, CMS, QW and BC; visualization, CMS and QW.}

\funding{CMS is funded by Comisión Nacional de Investigación Ciencia y Tecnología (CONICYT, currently ANID) through Programa Formacion de Capital Avanzado (PFCHA), Doctoral scholarship Becas Chile: CONICYT PFCHA/DOCTORADO BECAS CHILE/2016 - 72170507. QW was supported by AFOSR grant FA2386-18-1-4028. We also acknowledge the grant: Categorical Theories of Consciousness: Bridging Neuroscience and Fundamental Physics, FQXi-RFP-CPW-2018.}

\acknowledgments{The authors appreciate valuable feedback and discussions from Sean Tull, as well as very constructive comments from anonymous reviewers.}

\conflictsofinterest{The authors declare no conflict of interest.} 




\reftitle{References}






\externalbibliography{yes}
\bibliography{library.bib}



\end{document}